# Genetic Co-Occurrence Network Across Sequenced Microbes

Pan-Jun Kim[1,2], Nathan D. Price[1,3,4,5]*

1 Institute for Genomic Biology, University of Illinois, Urbana, Illinois, United States of America
2 Present addresses: Asia Pacific Center for Theoretical Physics, Pohang, Gyeongbuk, Republic of Korea; Department of Physics, POSTECH, Pohang, Gyeongbuk, Republic of Korea
3 Department of Chemical and Biomolecular Engineering, University of Illinois, Urbana, Illinois, United States of America
4 Center for Biophysics and Computational Biology, University of Illinois, Urbana, Illinois, United States of America
5 Present address: Institute for Systems Biology, Seattle, Washington, United States of America

* E-mail: nprice@systemsbiology.org (NDP)

## Abstract

The phenotype of any organism on earth is, in large part, the consequence of interplay between numerous gene products encoded in the genome, and such interplay between gene products affects the evolutionary fate of the genome itself through the resulting phenotype. In this regard, contemporary genomes can be used as molecular records that reveal associations of various genes working in their natural lifestyles. By analyzing thousands of orthologs across ~600 bacterial species, we constructed a map of gene-gene co-occurrence across much of the sequenced biome. If genes preferentially co-occur in the same organisms, they were called herein *correlogs*; in the opposite case, called *anti-correlogs*. To quantify correlogy and anti-correlogy, we alleviated the contribution of indirect correlations between genes by adapting ideas developed for reverse engineering of transcriptional regulatory networks. Resultant correlogous associations are highly enriched for physically interacting proteins and for co-expressed transcripts, clearly differentiating a subgroup of functionally-obligatory protein interactions from conditional or transient interactions. Other biochemical and phylogenetic properties were also found to be reflected in correlogous and anti-correlogous relationships. Additionally, our study elucidates the global organization of the gene association map, in which various modules of correlogous genes are strikingly interconnected by anti-correlogous crosstalk between the modules. We then demonstrate the effectiveness of such associations along different domains of life and environmental microbial communities. These phylogenetic profiling approaches infer functional coupling of genes regardless of mechanistic details, and may be useful to guide exogenous gene import in synthetic biology.

# Introduction

An important challenge in biology is to understand the relationships between an organism's genotype and its phenotype, involving dissection of myriad interdependencies among various cellular components, such as genes, proteins, and small molecules [1,2]. Recent efforts to map protein-protein interactions [3,4], together with efforts to reconstruct metabolic and regulatory networks at the genome scale [5,6,7], offer promising opportunities to investigate emergent biological phenomena of interacting biomolecules inside cells. Complex interdependencies among the products of individual genes do not necessarily imply molecular interactions by direct physical contacts but also include more macroscopic associations exhibited at e.g. biochemical pathway levels, as has been claimed in epistasis research [8]. These genetic interdependencies, regardless of their underlying mechanisms, may leave a trace on the composition of genomes through evolutionary processes.

One important set of analyses that have been used successfully over the past decade is based on phylogenetic profiles as introduced in [9]. Here, we analyze a type of phylogenetic profile − the patterns of the presence or absence of orthologs across many organisms − to find genes with favored co-occurrence in the same genomes (called herein *correlogs*) or disfavored co-occurrence (called *anti-correlogs*) suggesting their putative functional coupling. Such analysis of correlogy and anti-correlogy can help uncover global gene associations conserved at the biome level, beyond those specific to any particular organism. This information is distinct from genetic interactions inferred from, for example, double-mutant data [10,11], which relate to gene relationships within the specific organisms employed in the experiments. In particular, knowledge of the association of genes not coexisting inside considered specific organisms can be applied for heterologous gene expression in synthetic biology [12], and interest in anti-correlogy itself inevitably tends to target heterologous genes. It should also be noted that fitness of an organism in its natural habitat does not necessarily coincide with fitness in a laboratory [13], and ortholog profiles of organisms may thus encompass genetic relationships in environmental and ecological contexts not readily captured by laboratory experiments.

The importance of co-occurring orthologs as a means to gain insight into gene relationships is well appreciated in many previous studies [9,14,15,16,17,18,19]; however, these studies have largely focused on improving the identification of molecular-level interactions rather than on systematic analysis of the global organization of gene associations, as is the focus herein. In addition, we have included the removals of indirect gene-gene correlations in analyzing co-occurrence patterns to reduce false positive associations. We here take a comprehensive and systematic approach to study from a global perspective the biome-wide associations of genes as manifest through patterns of correlogy and anti-correlogy across sequenced bacteria.

# Results

**Characterization of correlogy and anti-correlogy**

We surveyed the presence or absence of gene orthologs across 588 different bacterial species (Table S1) on the basis of orthology data available from the Kyoto Encyclopedia of Genes and Genomes (KEGG) [20]. Beyond simply measuring co-occurrence of these genes, we evaluated degrees of correlogy or anti-correlogy between the genes within the context of direct associations in biological activities, using methods to help avoid vertical co-inheritance effects, transitivity effects, and other spurious correlations. In particular, we attempted to reduce the contribution of indirect (anti-)correlogous relationships that result from transitivity effects: if genes $i$ and $j$ are each correlated with third common genes in terms of the presence or absence across species, genes $i$ and $j$ can also appear to be correlated with each other even in the absence of direct association between them. Therefore, simple correlations calculated from the co-occurrence pattern can suffer from these indirect correlations. Such removal of indirect correlations in this manner has not generally been taken into account in previous studies on ortholog profiles [9,14,15,16,17,18,19], or also with similar types of correlation calculations in other fields to infer disease comorbidity networks [21,22,23] or social networks [24]. Nevertheless, filtering out these transitivity effects has been of critical importance in e.g. reverse engineering of transcriptional regulatory networks [25,26], and we employed such ideas to reduce transitivity effects when quantifying correlogy and anti-correlogy. As a result, for every pair of a total of 2085 genes, we assigned $w_{ij}$ of which magnitude increase away from zero measures the magnitude of correlogy ($w_{ij} > 0$) or anti-correlogy ($w_{ij} < 0$) between genes $i$ and $j$ in a pair (see Methods).

**Biochemical and phylogenetic properties**

It is worthwhile to address how much correlogous relationships overlap with molecular-level interactions. To test this idea, we compared the distribution of $w_{ij}$ for physically-binding proteins in *Escherichia coli* with that for arbitrary pairs of the proteins [27], and found that highly correlogous proteins are much likely to physically interact (Figure 1A). Indeed, the average of $w_{ij}$ for directly-binding proteins was 6.8 times larger than that for the second nearest proteins in the protein interaction network ($P = 7.1 \times 10^{-55}$), and 10.5 times larger than that for all the protein pairs ($P = 8.8 \times 10^{-57}$; Figure 1B and Methods). Instead of $w_{ij}$, if we use the simple co-occurrence measure $r_{ij}$ that precedes $w_{ij}$ before the alleviation of transitivity effects (Methods), the overall distribution of $r_{ij}$ is heavily biased toward positive $r_{ij}$'s (Figure 1C), and the enrichment of high $r_{ij}$'s in physically-binding protein pairs is relatively weak ($P = 4.0 \times 10^{-8}$ against the second nearest proteins) although still significant. Moreover, the network-topological signature appearing for $w_{ij}$ in Figure 1B becomes very distorted for $r_{ij}$ with a hump at the tenth nearest proteins in Figure 1D. Note that a hump at the tenth nearest proteins represents average $r_{ij}$ larger than averages at nearer proteins, possibly contributed to by indirect correlations from transitivity but unlikely to be biologically meaningful. Thus, the elimination of effects caused by transitivity is critical to identifying biologically meaningful correlogous and anti-correlogous relationships between genes. Another simple co-occurrence measure we tried, the mutual information $I_{ij}$ of genes $i$ and $j$ (Methods), reflects better the physical interactions than $r_{ij}$ but still less than $w_{ij}$ ($P = 3.4 \times 10^{-28}$ against the second nearest proteins; Figures 1E and 1F). Again, $I_{ij}$ leaves a small hump at the tenth nearest proteins in Figure 1F, and by its definition does not directly provide positive or negative signs of gene relationships themselves which are important in our study.

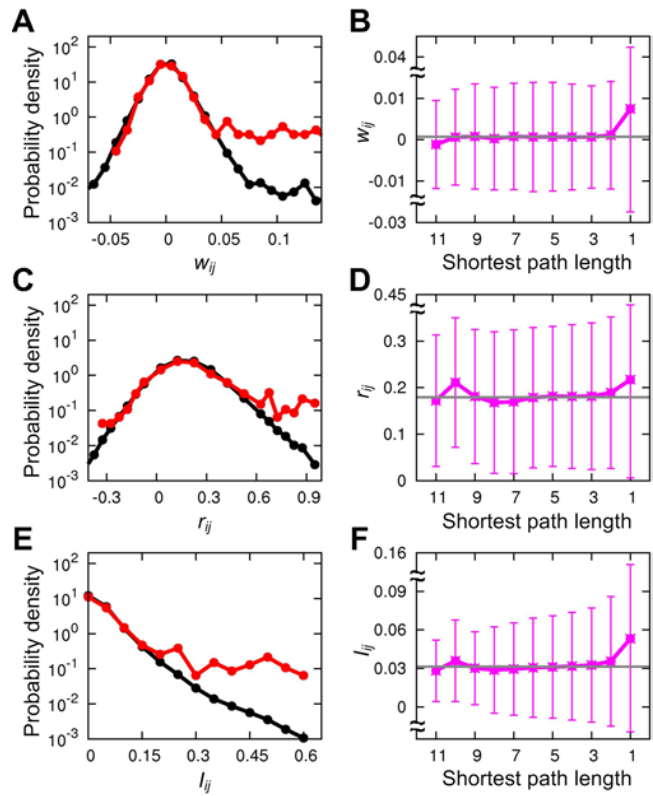

**Figure 1. Biochemical properties of correlogous and anti-correlogous gene associations.** (A) The probability density of $w_{ij}$ for physically-interacting protein pairs (red) and that for all protein pairs (black) in the *E. coli* protein interaction network. The probability density means the fraction of protein pairs in a unit interval of $w_{ij}$. Note that the red line is not only skewed into $w_{ij} > 0$ but is also rapidly truncated at $w_{ij} < 0$ compared with the black line, indicating that physically-interacting proteins are highly enriched for $w_{ij} > 0$ compared with arbitrary pairs of proteins. (B) The average $w_{ij}$ for each value of the shortest path length among pairs of proteins in the protein interaction network. (C and D) Plotted for $r_{ij}$ in the same ways as (A) and (B). (E and F) Plotted for $I_{ij}$ in the same ways as (A) and (B). Gray horizontal lines in (B), (D), and (F) indicate the global average, and error bars represent standard deviations.

The data in Figure 1A suggest a natural boundary for separating a regime enriched with functionally-obligatory protein interactions ($w_{ij} > 0.045$) from that with conditional or transient interactions [28,29]. Specifically, the probability density of $w_{ij}$ for physically-binding proteins deviates strongly from that for arbitrary pairs of proteins at $w_{ij} > 0.045$, but almost overlaps with that for the arbitrary pairs at the rest $w_{ij}$ (except for the absence of strong anti-correlogy). In the former regime of $w_{ij}$, there thus might be present functionally-obligatory relationships between physically interacting proteins to constrain them with large correlogous associations, as indicated by the rich presence of operonic genes whose transcriptions are precisely co-regulated in time (Text S1 and Figure S1). Even if we exclude these operonic gene pairs, the transcripts of the gene pairs with $w_{ij} > 0.045$ still tend to be more co-expressed than the others ($P = 0.03$), implying more obligatory interactions between them (Text S1 and Figure S1). Taken together, integration of protein-protein interaction data with the heterogeneous data of correlogous relationships provides a powerful means for identifying potentially functionally-obligatory interactions conserved across evolution from mere binding events.

It is of course expected that other forms of molecular interactions than physically-binding interactions would also contribute to correlogy and anti-correlogy. Table 1 presents some cases of the highest $|w_{ij}|$'s; *rfbF* encodes the enzyme to catalyze the reaction, CTP + α-D-glucose 1-phosphate → diphosphate + CDP-glucose, and the produced CDP-glucose is subsequently converted by the enzyme from the correlogous gene, *rfbG*, into CDP-4-dehydro-6-deoxy-D-glucose. The correlogous relationship between *rfbF* and *rfbG* thus appears to be a consequence of the need for the second reaction to proceed when the first occurs in order to achieve the relevant biological functions. On the other hand, two $NAD^+$ synthases, one utilizing ammonia as an amide donor to produce $NAD^+$ and the other utilizing glutamine as an amide donor instead, are highly anti-correlogous to each other, and the anti-correlogy reflects that both processes operating simultaneously in the same organism have been consistently selected against. This fact might be related to the distinct modes of regulating cellular nitrogen inside bacteria, as pleiotropic nitrogen utilization mutations can be found at $NAD^+$ synthases [30]. In general, genes associated with similar functions were enriched with correlogous and anti-correlogous relationships (Figure S2 and Text S1). Indeed, the average of $w_{ij} > 0$ for isozymes based on the same Enzyme Commission numbers was 5.8 times larger than that for arbitrary pairs of enzymes ($Z = 83.69$), and the average of $w_{ij} < 0$ for the isozymes was 2.9 times larger than that for arbitrary enzyme pairs ($Z = 24.11$). Thus, isozymes exhibit high levels of correlogy and anti-correlogy compared to arbitrary enzyme pairs, although the effect is more pronounced for the correlogy side.

**Table 1. Three most correlogous or anti-correlogous pairs of genes.**

| $w_{ij}$ | KEGG identifier | Name | Description |
|---|---|---|---|
| | | *Correlogous* | |
| 0.4037 | K05878 | *dhaK* | Dihydroxyacetone kinase, N-terminal domain |
| | K05879 | *dhaL* | Dihydroxyacetone kinase, C-terminal domain |
| 0.3982 | K00978 | *rfbF* | Glucose-1-phosphate cytidylyltransferase |
| | K01709 | *rfbG* | CDP-glucose 4,6-dehydratase |
| 0.3975 | K01977 | 16S rRNA, *rrs* | 16S ribosomal RNA |
| | K01980 | 23S rRNA, *rrl* | 23S ribosomal RNA |
| | | *Anti-correlogous* | |
| −0.2188 | K03785 | *aroD* | 3-dehydroquinate dehydratase I |
| | K03786 | *aroQ, qutE* | 3-dehydroquinate dehydratase II |
| −0.2107 | K01916 | *nadE* | NAD$^+$ synthase |
| | K01950 | *NADSYN1, QNS1, nadE* | NAD$^+$ synthase (glutamine-hydrolysing) |
| −0.1817 | K00756 | *pdp* | Pyrimidine-nucleoside phosphorylase |
| | K00758 | *deoA* | Thymidine phosphorylase |

To evaluate the overall correlogous couplings around individual genes, we defined for a given gene $i$, $S_i^p = \sum_j^{w_{ij}>0} w_{ij}$, where the summation was taken over all other gene $j$'s satisfying $w_{ij} > 0$. Likewise, we can define $S_i^n = \sum_j^{w_{ij}<0} |w_{ij}|$. $S_i^p$ and $S_i^n$ quantify how tightly gene $i$ is associated to other genes through correlogy and anti-correlogy, respectively. Having both the largest $S_i^p$ and $S_i^n$ was gene *rpmJ*, which encodes ribosomal protein L36. The next largest were $S_i^p$ and $S_i^n$ of DNA-modifying genes, while the smallest ones were owned by flagella-related genes (Table S2). In general, $S_i^p$ and $S_i^n$ of each gene $i$ are very positively correlated (Text S1 and Figure S3).

How broadly genes are phylogenetically distributed would affect or be affected by the strength of interactions with other genes. Therefore, the degree of phylogenetic spread of any given gene is expected to be correlated with its $S_i^p$ and $S_i^n$. Rather surprisingly, our analysis suggests that species-level spread of genes does not exhibit such correlations (Figure 2A), but spread at a higher taxonomic level – at the phylum level – reveals clear correlations with $S_i^p$ and $S_i^n$; as genes inhabit diverse phyla, their $S_i^p$ and $S_i^n$ tend to increase continuously until saturated at a plateau ($Z = 79.94$; Figure 2B and Methods). Since different phyla (phylogenetically distant) may represent disparate cell types more clearly than different species (phylogenetically close), our result suggests that how many such disparate cellular conditions genes inhabit at the phylum level could substantially evolve or be influenced by the strength of the genetic interdependencies around the genes.

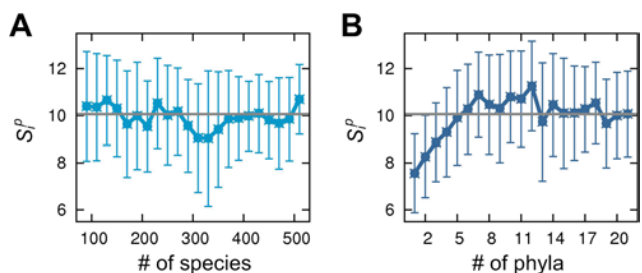

**Figure 2. Correlation with phylogenetic dispersion.** (A) The average $S_i^p$ of genes for each number of species where the genes are present, and (B) that for each number of phyla where the genes are present. For (A) and (B), $S_i^n$ also behaves similarly to $S_i^p$ (Text S1 and Figure S3). Gray horizontal lines indicate the global average, and error bars represent standard deviations.

## Global organizational properties

To systematically view the essential structure of correlogous and anti-correlogous relationships from a global architectural perspective, we constructed a maximum relatedness subnetwork (MRS) [24], in which each gene *i* points to two other genes *j* and *j'* by different categories of edges that represent the most correlogous ($\max_j w_{ij} > 0$) and anti-correlogous ($\min_{j'} w_{ij'} < 0$) genes to gene *i*, respectively (Figures 3A-3C and Table S3). By definition, genes in MRS are not necessarily reciprocally linked. As such, one can easily recognize that following a series of genes in either of correlogy or anti-correlogy direction of links, the magnitude of $w_{ij}$ of each link increases until encountering reciprocal links. MRS provides the 'backbone' structure of a considered network, and has been demonstrated as particularly useful for detecting modular structures of complex networks [24]. One example is for two isocitrate dehydrogenases, *IDH1* and *IDH3*, which encode similar enzymes depending on $NADP^+$ and $NAD^+$, respectively, and are anti-correlogously associated in the MRS. Consistent with our observations, their distinct phylogenetic profiles have been discussed in previous studies [31]. Also in the MRS, *IDH1* is correlogously associated with *gltA* that encodes citrate synthase, and the enzyme activities from *IDH1* and *gltA* are indeed known as elaborately coordinated in a cell for efficient growth on acetate [32]. For another example, β-glucosidase (*bglB*), which breaks down cellobiose into β-D-glucose and can be industrially useful for lignocellulose conversion for biofuel production [33], is correlogously associated in the MRS with L-rhamnose mutarotase (*rhaM*). We expect that this mutarotase may convert the product of β-glucosidase into α-D-glucose if the host cell prefers the α form to the β form, and thus expressing both the two genes, *bglB* and *rhaM*, may introduce a synergetic effect in cellobiose metabolic engineering. Furthermore, Figure 3C illustrates that, along correlogy direction, the MRS arranges sequentially xylose transport protein (*xylF*), xylose isomerase (*xylA*), and xylulokinase (*xylB*), the same order as their arrangement in the xylose metabolism pathway (Figure 3D). It is interesting to note that the MRS brings relatively less characterized genes such as *xylF* to attention by informing of their strong correlogous relationships with other genes. As such, Figure 3C does specify the identities of genes important for xylose metabolic engineering [34] that may benefit from simultaneous heterologous expression of *xylF*, *xylA*, and *xylB*. Also, by their link directionality the MRS provides the information that *xylA* and *xylB* are more associated than *xylA* and *xylF*.

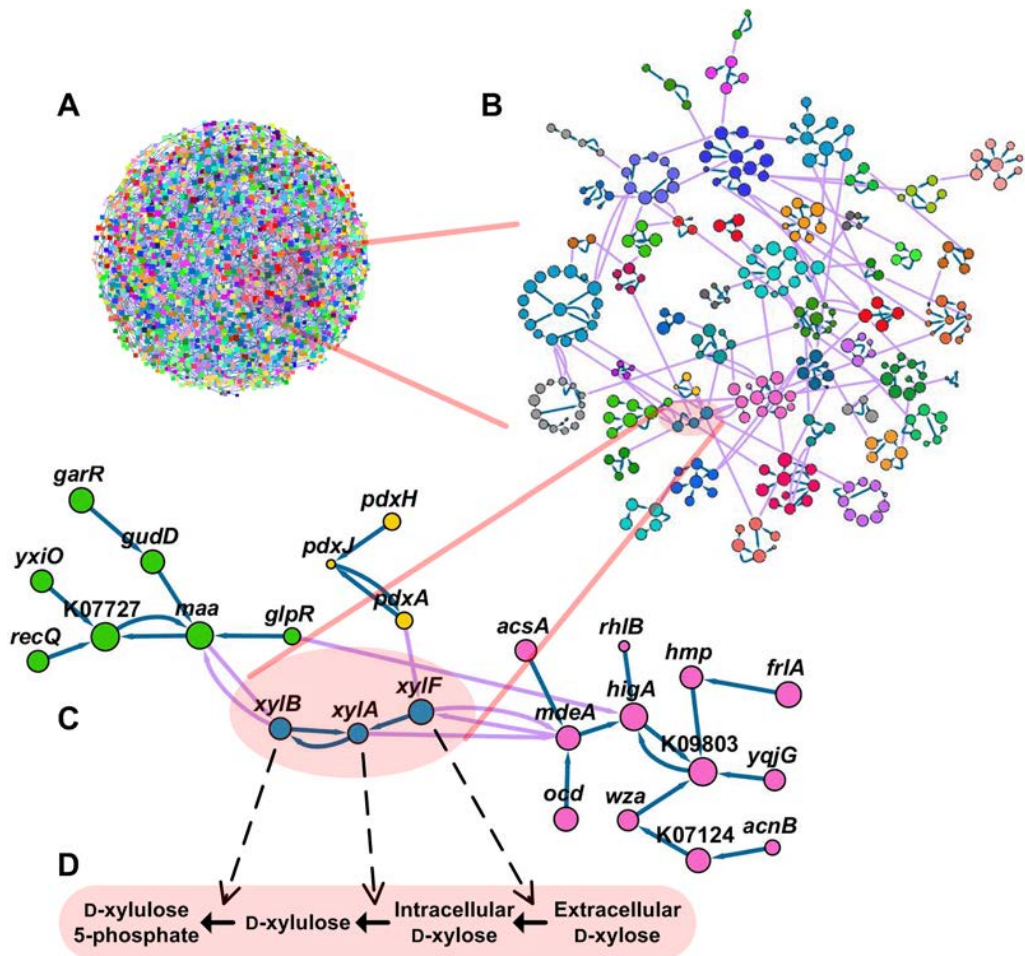

**Figure 3. MRS of gene associations.** (A−C) Large-scale to small-scale overviews of the MRS. A part of the whole MRS in (A) is magnified in (B) to reveal its clear modular structure. Circles represent genes, and each gene *i* is arrowed to gene *j* having the largest $|w_{ij}|$ with $w_{ij} > 0$ (dark arrow) or $w_{ij} < 0$ (light arrow). Size of each circle corresponds to $S_i^p$ of the gene, and circles are colored the same if linked via $w_{ij} > 0$. In (B) and (C), different groups of genes linked via $w_{ij} > 0$ are shown to be seamlessly connected by links of $w_{ij} < 0$. In (C), gene names or KEGG identifiers appear for relevant circles. (D) Transport and biochemical conversions of xylose catalyzed by the proteins from genes in (C).

We found that all genes in the MRS are decomposed into 483 different small subgroups (Table S3), of which each includes correlogously associated genes yet not linked correlogously to any genes in the other subgroups. Nevertheless, the vast majority (99.5%) of anti-correlogy links bridge the gaps between different modules in the MRS, binding all of them as a single giant component (Figures 3B and 3C). Thus, the MRS results in clear modular structure among correlogs, with these modules interrelated through anti-correlogous relationships. Different genes in each correlog group are highly likely to perform biological tasks in a common functional category ($Z = 43.16$ for KEGG categories; see Methods). For every pair of possible functional categories, we also analyzed how likely genes of the functional categories in a pair would belong together to the same correlog groups (Methods). As expected, genes of the same KEGG functional categories tend to belong to the same correlog groups, but deviations from this behavior are also informative (Figure 4A and Table S4). For example, genes of functional category *Folding, sorting and degradation* and those of another category *Metabolism of other amino acids* significantly tend to be affiliated to the same correlog groups ($P < 10^{-4}$). The former category involves a number of molecular chaperones and RNA helicases, while the latter involves enzymes to synthesize glutathione and D-glutamate. Since glutathione serves as the major endogenous antioxidant and D-glutamate decorates bacterial cell walls, our results support the previous observation that the corresponding enzymes are likely to be actively in concert with chaperones and RNA helicases when subject to oxidative or cell-wall stress [35,36]. Likewise, genes of functional category *Xenobiotics biodegradation and metabolism* and those of category *Folding, sorting and degradation* are highly likely to be together in the same correlog groups (Figure 4A and Table S4), which can be understood in the similar context of cellular stress response induced by xenobiotic compounds such as benzoate and bisphenol [37].

We found that the number of genes in each correlog group approximately follows an exponential distribution (Figure 4B). We expect that each correlog group may serve as a toolbox for importing exogenous genes and functions into a cell, as *xylF*, *xylA*, and *xylB* above form a single correlog group themselves (Figures 3C and 3D). Furthermore, the prevalence of anti-correlogy links between correlog groups as shown in Figures 3B and 3C extends the concept of anti-correlogy from single genes to correlog groups and allows for evaluating the anti-correlogous associations between, rather than within, the correlog groups (Table S5 and Text S1). For example, a correlog group containing aldehyde-related dehydrogenases was anti-correlogously associated with another group containing *perR*, peroxide stress response regulator, and this anti-correlogy between the two groups might be involved in a problem that arises in controlling cellular redox state [38].

If a correlog group represents a repertoire of genes that tend to coexist in the same organisms, genes in individual organisms when mapped to MRS should be densely distributed around the correlog groups rather than distributed uniformly. This hypothesis can be straightforwardly checked by enumerating the number of different correlog groups that the genes are mapped to, and comparing this value with the number of different correlog groups harboring the same number of genes but mapped randomly. If the former is smaller than the latter, the genes can be regarded as more clustered around correlog groups than by chance. As would be expected, genes from each bacterial species show much more clustered behaviors than by chance ($-13.20 \leq Z \leq -3.68$ for all bacteria; Figures 4C-4E and Methods). To the MRS, we can also map orthologs in each species from the other superkingdoms, archaea and eukaryotes. Interestingly, although the MRS in this study is based on bacterial data, orthologs from archaea and eukaryotes show somewhat weakened but still significantly clustered behaviors around the correlog groups, reflecting a significant conservation of gene associations across different domains of life [$-5.97 \leq Z \leq -2.64$ for archaea except for one species with $Z = -1.32$ (Figure 4C) and $-6.47 \leq Z \leq -2.23$ for eukaryotes except for five species with $-1.93 \leq Z \leq -1.39$ (Figure 4D); see Methods]. Among these observations, less significantly clustered behaviors come from the species having small numbers of genes mapped to the MRS, and this result might be due to insufficient mappings or minimally-required diversity of genes for any biological systems. On the other hand, considering that correlogs tend to coexist in individual organisms, it might be challenging to examine the above clustering issues for environmental microbial communities [39], as each of the environmental samples typically contains a number of different species. Again, we mapped to the MRS the genes from twelve diverse environmental sources such as human and mouse guts, deep-sea whale fall carcasses, and uranium contaminated ground water [40], and found still significant clustering of those genes around the correlog groups ($-9.26 \leq Z \leq -2.10$ except for one sample with less significant $Z = -0.95$; Figure 4E and Methods). Accordingly, this result encourages us to even conjecture the identities of undetected but existing genes in environmental samples based on those of detected genes by applying the knowledge of correlogous gene associations. It would also be interesting to identify correlog groups harboring cosmopolitan genes [41] in a given environment, as these groups can represent together environment-specific genetic contents rather than species-specific ones.

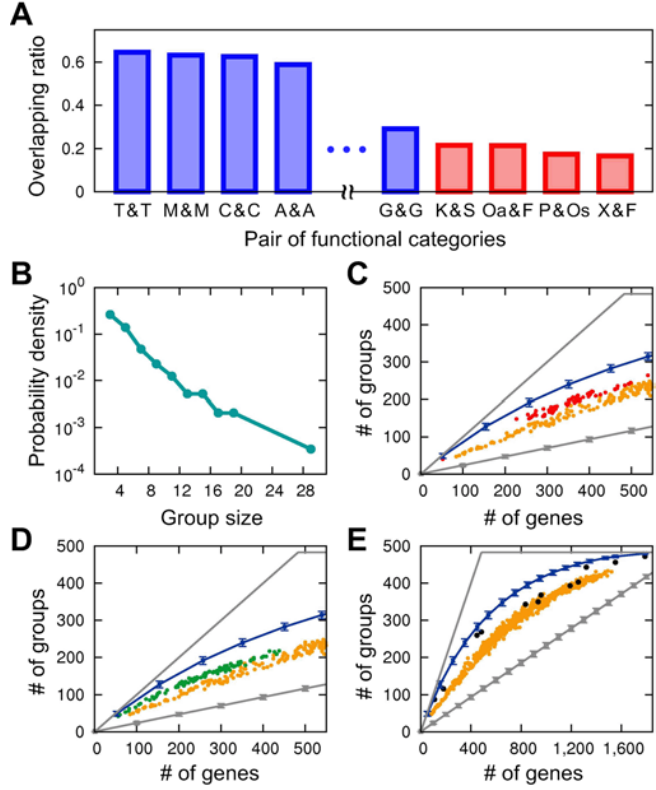

**Figure 4. Characteristics of correlog groups in the MRS.** (A) Plotted in descending order is how likely genes in a given pair of functional categories belong to the same correlog groups (plotted for $P < 10^{-3}$). Data of intermediate values are omitted for visual clarity, and all of them can be found in Table S4. Blue is for the same functional categories in a pair, and red is for different functional categories in a pair. Functional categories are abbreviated as the followings. T: Membrane transport; M: Cell motility; C: Carbohydrate metabolism; A: Amino acid metabolism; G: Glycan biosynthesis and metabolism; K: Kinase and peptidase; S: Signal transduction; Oa: Metabolism of other amino acids; F: Folding, sorting and degradation; P: Biosynthesis of polyketides and terpenoids; Os: Biosynthesis of other secondary metabolites; X: Xenobiotics biodegradation and metabolism. (B) The probability density of the number of genes in each correlog group. (C−E) Clustering of genes from individual species and environmental samples. For each bacterial species, the number of genes mapped to the MRS (*x* axis) and the number of different correlog groups including the genes (*y* axis) are plotted in yellow. In the same way, for each archaeal species, plotted in red in (C), for each eukaryotic species, plotted in green in (D), and for each environmental sample, plotted in black in (E). In (C)-(E), for comparison with the null model, the number of different correlog groups including randomly mapped genes is plotted in blue with the error bars representing standard deviations. Although for some data points the error bars in blue look overlapped with the symbols in red and green, this is simply because of the large size of the symbols used to facilitate visual examination. Gray lines at the upper and lower sides in (C)-(E) denote the number of different correlog groups with the most sparsely-distributed mapped genes and the maximally-crowded mapped genes, respectively, while the latter allows for multiple configurations of gene mappings having standard deviations denoted by error bars. In other words, the gray lines set the absolute limits of the number of different correlog groups for each number of genes. Most data points in yellow, red, green, and black are located far below blue error bars, illustrating significant clustering of genes around correlog groups across different domains of life and environmental microbial communities.

## Discussion

Recent advances in understanding genome-scale intracellular networks have been enabled by the availability of high-throughput experimental techniques that permit network data to be collected on a scale far larger than previously possible [1,2]. These networks help capture and quantify the functional interplay amongst genes that drives essentially all cellular phenotypes in the environment. Correlogy and anti-correlogy can be regarded as the natural outcome of such ubiquitous interactions between genes, molded through a long evolutionary process that tests beneficial effects of numerous gene combinations for a cell. The resulting 'sociology' of genes provides rich implications in rapidly growing fields such as function prediction of uncharacterized genes [27], gap-filling in genome-scale biochemical model reconstruction [42], and synthetic biology [12]. In synthetic biology, information on both correlogs and anti-correlogs may be useful to screen additional gene candidates to be expressed or silenced contingent on primary genes of interest, and such considerations should inform attempts to construct optimal recombinant strains. Intriguingly, the information on correlogy and anti-correlogy can be uncovered even when the underlying mechanisms of these gene associations are unknown. Finally, future extension of our work towards archaea and eukaryotes will allow comparative analysis of correlogy and anti-correlogy for different domains of life, offering a fascinating opportunity of understanding evolution of genetic associations.

## Acknowledgments

We thank Nigel Goldenfeld, Carl Woese, Yong-Su Jin, Nicholas Chia, Charu Gupta Kumar, Suk-Jin Ha, and Jin-Ho Choi for critical discussions and Jaeyun Sung, Younhee Ko, and James Eddy for technical support.

## Methods

All methods used in this study, including the methods of various data analysis, are provided in Text S1 with a detailed description. Below, we present an abbreviated version that describes the essence of our analysis.

**Quantification of correlogy and anti-correlogy**

We downloaded orthology data from the KEGG database [20], and surveyed the presence or absence of each ortholog across different bacterial species (see Table S1 for the full list of the bacterial species considered here). In order to capture functional interactions of genes reflected in their co-occurrence patterns across species, we take into account the genes (i.e., orthologs) not too lowly nor too highly prevalent across species; in the case of too lowly (highly) prevalent genes, there do not exist so many species with (without) the genes, making it hard to judge whether these few co-presences (co-absences) of the genes actually come from their functional interactions. In other words, without filtering, spurious correlations from non-functional origins may emerge, simply by vertical co-inheritance of genes or by chance. Specifically, if $E_i$ denotes the number of species containing gene $i$ and $N$ denotes the total number of species, one can define $X_i = \min(E_i, N-E_i)$ for each gene $i$. The probability density of $X_i$ approximately follows the power-law decay as long as $X_i \geq X_{th} = 80$, and we chose the genes with $X_i \geq X_{th}$ to prevent spurious correlations that could occur at low $X_i$ deviating from the power-law trend observed at large $X_i$.

To evaluate direct gene associations while alleviating transitivity effects in charge of indirect correlations between genes, we applied the partial correlation method employed in graphical Gaussian models [26], of which superiority over many other methods was demonstrated in reverse engineering of transcriptional regulatory networks [25]. To implement this method, we start with calculating the Pearson correlation $r_{ij}$ for binary variables of presence and absence of genes $i$ and $j$:

$$r_{ij} = \frac{C_{ij}N - E_i E_j}{\sqrt{E_i E_j (N - E_i)(N - E_j)}},$$

where $C_{ij}$ is the number of species containing both genes $i$ and $j$. Next, to reduce indirect correlations between genes $i$ and $j$, we calculate the partial correlation $w_{ij}$ using $r_{ij}$:

$$w_{ij} = -\frac{p_{ij}}{\sqrt{p_{ii} p_{jj}}},$$

where $p_{ij}$ is the $(i, j)^{\text{th}}$ component of an inverse matrix of $r_{ij}$. However, in this study, the number of genes (= 2085) is much larger than the number of species (= 588), yielding an ill-conditioned problem for matrix $r_{ij}$. To overcome this problem, we applied the shrinkage estimation derived by Schäfer and Strimmer [26]. Specifically, Schäfer and Strimmer obtained a regularized estimator of $r_{ij}$ combining analytic determination of shrinkage intensity from the Ledoit-Wolf theorem [43]. The following is the resultant estimator $r^*_{ij}$ that simply substitutes for $r_{ij}$ in the above calculation of $w_{ij}$:

$$r^*_{ij} = \delta_{ij} + r_{ij}(1-\delta_{ij})\min[1,\max(0,1-\lambda)],$$

where $\delta_{ij}$ is the Kronecker delta symbol and $\lambda$ is given by

$$\lambda = \frac{\sum_{i \neq j}\left\langle\left[(x_{ki}-\langle x_{ki}\rangle_k)(x_{kj}-\langle x_{kj}\rangle_k)\bigg/\left(\frac{N-1}{N}\sigma_i\sigma_j\right)-r_{ij}\right]^2\right\rangle_k}{(N-1)\times\sum_{i \neq j}r_{ij}^2}.$$

Here $x_{ki} = 1$ if gene $i$ is present in species $k$, otherwise, $x_{ki} = 0$, $\langle\cdots\rangle_k$ denotes the average over species $k$'s, and $\sigma_i$ is the standard deviation of $x_{ki}$ over species $k$'s. As a result, the obtained $w_{ij}$ was used to quantify correlogy ($w_{ij} > 0$) or anti-correlogy ($w_{ij} < 0$) between genes $i$ and $j$. For comparative analysis, the mutual information $I_{ij}$ of genes $i$ and $j$ between $\{x_{ki}\}$ and $\{x_{kj}\}$ across species $k$'s [14] was also calculated.

**Significance analysis of correlation between $w_{ij}$ (or $r_{ij}$, $I_{ij}$) and protein interaction**

We calculated the average of $w_{ij}$ ($w^{ppi}$) from the pairs of physically-binding proteins in *E. coli* [27], and obtained its $P$ value by generating the distribution of average $w_{ij}$ ($w^{null}$) from the same number of, but arbitrarily-mated pairs of the proteins as distant as the given shortest path length in the protein interaction network. The central limit theorem ensured that this null distribution converged well to the Gaussian distribution, providing the $P$ value for how frequently $w^{null}$ exceeds $w^{ppi}$. Similar analyses were also performed for $r_{ij}$ and $I_{ij}$.

**Evaluation of $S_i^p$ and $S_i^n$**

In order to characterize how tightly each gene $i$ is correlogously associated to other genes, we calculated $S_i^p = \sum_j^{w_{ij}>0} w_{ij}$, where the summation was taken over all other gene $j$'s satisfying $w_{ij} > 0$. In a similar way, we calculated $S_i^n = \sum_j^{w_{ij}<0} |w_{ij}|$ for anti-correlogous couplings around gene $i$.

**Significance analysis of correlation between $S_i^p$ and phylum-level dispersion**

Let $n^{phyla}$ be the number of different phyla where genes are present. For genes with $n^{phyla} < 7$ (Figure 2B), we obtained the slope of $S_i^p$ against $n^{phyla}$ by linear regression, and normalized it by multiplying $\langle n^{phyla}\rangle/\langle S_i^p\rangle$. From surrogate data with randomly-permuted gene presences across species, we also generated an ensemble of such normalized slopes for $n^{phyla} < 7$, and calculated the Z score of the actual value.

**Characterization of the maximum relatedness subnetwork (MRS)**

For any given weighted network, one can simplify its structure by constructing the MRS composed only of highly weighted edges [24]. Specifically, in the MRS of this study, each gene $i$ points to only two genes $j$ and $j'$ by different categories of edges that represent the most correlogous ($\max_j w_{ij} > 0$) and anti-correlogous ($\min_{j'} w_{ij'} < 0$) genes to gene $i$, respectively. We found that all genes in the MRS here are decomposed into 483 different small subgroups, of which each includes correlogously associated genes yet not linked correlogously to any genes in the other subgroups. These subgroups in the MRS were termed correlog groups.

**Functional coherence of correlog groups in the MRS**

For given correlog group $g$ in the MRS and given functional category $c$ of genes, we can calculate $f_c^g = \tilde{N}_c^g/\tilde{N}^g$, where $\tilde{N}_c^g$ is the number of genes affiliated to both correlog group $g$ and functional category $c$, and $\tilde{N}^g$ is the total number of genes affiliated at least to one functional category in correlog group $g$. Therefore, $f_c^g$ represents the uniformity of gene functions in a correlog group. Majority (57.1%) of correlog groups with $\tilde{N}^g > 1$ were shown to have at least one functional category $c$ satisfying $f_c^g = 1$ in each $g$. To calculate the corresponding Z score, we generated an ensemble of correlog groups with $\tilde{N}^g > 1$ by randomly exchanging genes of the same number of the affiliated functional categories.

For a given pair of functional categories $c1$ and $c2$, we can also define their overlapping ratio (Figure 4A and Table S4) as:

$$Y_{c1,c2} = \frac{\sum_g^{\tilde{N}^g>1} H\left(\sum_{i,j}^{i \neq j} a_{i,c1} a_{j,c2} d_{i,g} d_{j,g}\right)}{\sum_g^{\tilde{N}^g>1} H\left[\sum_i (a_{i,c1} + a_{i,c2}) d_{i,g}\right]},$$

where $i$ and $j$ are indices of genes, $a_{i,c}$ is 1 if gene $i$ belongs to functional category $c$, otherwise 0, $d_{i,g}$ is 1 if gene $i$ belongs to correlog group $g$, otherwise 0, and $H(x)$ is 1 if $x > 0$, otherwise 0. $Y_{c1,c2}$ quantifies how likely genes in $c1$ and $c2$ belong to the same correlog groups ($0 \leq Y_{c1,c2} \leq 1$). The corresponding P value was obtained by generating the null distribution in the same way as in the case of $f_c^g$ above.

**Effectiveness of correlog groups for different domains of life and environmental samples**

For each species or environmental sample, we counted the number ($n$) of correlog groups harboring the genes mapped to the MRS. We also obtained the mean ($\eta$) and the standard deviation ($\sigma$) of such numbers of correlog groups when the same number of genes are randomly mapped to the MRS (Figure 4C-4E):

$$\eta = \bar{n} - \sum_g \left(1 - \frac{N}{\sum_g N_g}\right)^{N_g}, \quad \sigma = \sqrt{\sum_g \left[\left(1 - \frac{N}{\sum_g N_g}\right)^{N_g} - \left(1 - \frac{N}{\sum_g N_g}\right)^{2N_g}\right]},$$

where $N$ is the number of genes mapped to the MRS, $g$ is the index of each correlog group, $N_g$ is the total number of genes in correlog group $g$, and $\bar{n}$ is the total number of correlog groups in the MRS. Accordingly, we calculate the $Z$ score of $n$ [$Z = (n-\eta)/\sigma$].

# References

1. Barabási A-L, Oltvai ZN (2004) Network biology: Understanding the cell's functional organization. Nat Rev Genet 5:101−113.
2. Oberhardt MA, Palsson BØ, Papin JA (2009) Applications of genome-scale metabolic reconstructions. Mol Syst Biol 5:320.
3. Jeong H, Mason SP, Barabási A-L, Oltvai ZN (2001) Lethality and centrality in protein networks. Nature 411:41−42.
4. Yu H, Braun P, Yıldırım MA, Lemmens I, Venkatesan K, et al. (2008) High-quality binary protein interaction map of the yeast interactome network. Science 322:104−110.
5. Covert MW, Knight EM, Reed JL, Herrgard MJ, Palsson BØ (2004) Integrating high-throughput and computational data elucidates bacterial networks. Nature 429:92−96.
6. Duarte NC, Becker SA, Jamshidi N, Thiele I, Mo ML, et al. (2007) Global reconstruction of the human metabolic network based on genomic and bibliomic data. Proc Natl Acad Sci U S A 104:1777−1782.
7. Chandrasekaran S, Price ND (2010) Probabilistic integrative modeling of genome-scale metabolic and regulatory networks in *Escherichia coli* and *Mycobacterium tuberculosis*. Proc Natl Acad Sci U S A 107:17845−17850.
8. Segrè D, DeLuna A, Church GM, Kishony R (2004) Modular epistasis in yeast metabolism. Nat Genet 37:77–83.
9. Pellegrini M, Marcotte EM, Thompson MJ, Eisenberg D, Yeates TO (1999) Assigning protein functions by comparative genome analysis: Protein phylogenetic profiles. Proc Natl Acad Sci U S A 96:4285−4288.
10. Costanzo M, Baryshnikova A, Bellay J, Kim Y, Spear ED, et al. (2010) The genetic landscape of a cell. Science 327:425−431.
11. Butland G, Babu1 M, Diaz-Mejia JJ, Bohdana F, Phanse S, et al. (2008) eSGA: *E. coli* synthetic genetic array analysis. Nat Methods 5:789−795.
12. Arkin AP (2008) Setting the standard in synthetic biology. Nat Biotechnol 26:771−774.
13. Rauch EM, Sayama H, Bar-Yam Y (2002) Relationship between measures of fitness and time scale in evolution. Phys Rev Lett 88:228101.
14. Huynen M, Snel B, Lathe W, III, Bork P (2000) Predicting protein function by genomic context: Quantitative evaluation and qualitative inferences. Genome Res 10:1204−1210.
15. Bowers PM, Cokus SJ, Eisenberg D, Yeates TO (2004) Use of logic relationships to decipher protein network organization. Science 306:2246−2249.
16. von Mering C, Jensen LJ, Kuhn M, Chaffron S, Doerks T, et al. (2007) STRING 7−recent developments in the integration and prediction of protein interactions. Nucleic Acids Res 35:D358−D362.

17. Jothi R, Przytycka TM, Aravind L (2007) Discovering functional linkages and uncharacterized cellular pathways using phylogenetic profile comparisons: A comprehensive assessment. BMC Bioinform 8:173.
18. Singh S, Wall DP (2008) Testing the accuracy of eukaryotic phylogenetic profiles for prediction of biological function. Evol Bioinform Online 4:217−223.
19. Ruano-Rubio V, Poch O, Thompson JD (2009) Comparison of eukaryotic phylogenetic profiling approaches using species tree aware methods. BMC Bioinform 10:383.
20. Kanehisa M, Goto S (2000) KEGG: Kyoto Encyclopedia of Genes and Genomes. Nucleic Acids Res 28:27−30.
21. Hidalgo CA, Blumm N, Barabási A-L, Christakis NA (2009) A dynamic network approach for the study of human phenotypes. PLoS Comput Biol 5: e1000353.
22. Lee D-S, Park J, Kay KA, Christakis NA, Oltvai ZN, et al. (2008) The implications of human metabolic network topology for disease comorbidity. Proc Natl Acad Sci U S A 105: 9880−9885.
23. Park J, Lee D-S, Christakis NA, Barabási A-L (2009) The impact of cellular networks on disease comorbidity. Mol Syst Biol 5:262.
24. Lee SH, Kim P-J, Ahn Y-Y, Jeong H (2010) Googling social interactions: Web search engine based social network construction. PLoS ONE 5:e11233.
25. Soranzo N, Bianconi G, Altafini C (2007) Comparing association network algorithms for reverse engineering of large-scale gene regulatory networks: Synthetic versus real data. Bioinformatics 23:1640−1647.
26. Schäfer J, Strimmer K (2005) A shrinkage approach to large-scale covariance matrix estimation and implications for functional genomics. Stat Appl Genet Mol Biol 4:32.
27. Hu P, Janga SC, Babu M, Díaz-Mejía JJ, Butland G, et al. (2009) Global functional atlas of *Escherichia coli* encompassing previously uncharacterized proteins. PLoS Biol **7**:e1000096.
28. Jones S, Thornton JM (1996) Principles of protein-protein interactions. Proc Natl Acad Sci U S A 93:13−20.
29. Han J-DJ, Bertin N, Hao T, Goldberg DS, Berriz GF, et al. (2004) Evidence for dynamically organized modularity in the yeast protein-protein interaction network. Nature 430:88−93.
30. Merrick MJ, Edwards RA (1995) Nitrogen control in bacteria. Microbiol Rev 59:604−622.
31. Zhu G, Golding GB, Dean AM (2005) The selective cause of an ancient adaptation. Science 307:1279−1282.
32. Walsh K, Schena M, Flint AJ, Koshland DE, Jr (1987) Compensatory regulation in metabolic pathways: Responses to increases and decreases in citrate synthase levels. Biochem Soc Symp 54:183−195.
33. Ha S-J, Galazka JM, Kim SR, Choi J-H, Yang X, et al. (2011) Engineered *Saccharomyces cerevisiae* capable of simultaneous cellobiose and xylose fermentation. Proc Natl Acad Sci U S A 108:504−509.
34. Brat D, Boles E, Wiedemann B (2009) Functional expression of a bacterial xylose isomerase in *Saccharomyces cerevisiae*. Appl Environ Microbiol 75:2304−2311.

35. Hossain MM, Nakamoto H (2003) Role for the cyanobacterial HtpG in protection from oxidative stress. Curr Microbiol 46:70−76.
36. Utaida S, Dunman PM, Macapagal D, Murphy E, Projan SJ, et al. (2003) Genome-wide transcriptional profiling of the response of *Staphylococcus aureus* to cell-wall-active antibiotics reveals a cell-wall-stress stimulon. Microbiology 149:2719−2732.
37. Ramos JL, Gallegos M-T, Marqués S, Ramos-González M-I, Espinosa-Urge M, et al. (2001) Responses of Gram-negative bacteria to certain environmental stressors. Curr Opin Microbiol 4:166−171.
38. Mongkolsuk S, Helmann JD (2002) Regulation of inducible peroxide stress responses. Mol Microbiol 45: 9−15.
39. Hugenholtz P, Tyson GW (2008) Microbiology: Metagenomics. Nature 455:481−483.
40. Markowitz VM, Chen I-MA, Palaniappan K, Chu K, Szeto E, et al. (2010) The integrated microbial genomes system: An expanding comparative analysis resource. Nucleic Acids Res 38:D382−D390.
41. Frigaard N-U, Martinez A, Mincer TJ, DeLong EF (2006) Proteorhodopsin lateral gene transfer between marine planktonic Bacteria and Archaea. Nature 439:847−850.
42. Kumar VS, Dasika MS, Maranas CD (2007) Optimization based automated curation of metabolic reconstructions. BMC Bioinformatics 8:212.
43. Ledoit O, Wolf M (2003) Improved estimation of the covariance matrix of stock returns with an application to portfolio selection. J Empir Finance 10: 603−621.